\documentclass[pre,twocolumn,groupedaddress,floatfix]{revtex4-1}
\usepackage{amssymb,amsmath}
\usepackage{graphicx}
\usepackage{subfigure}
\usepackage[english]{babel} 
\usepackage{float}
\usepackage{color}

\usepackage{lipsum}
\begin{document}
\newcommand{\be}{\begin{equation}}
\newcommand{\ee}{\end{equation}}
\newcommand{\rojo}[1]{\textcolor{red}{#1}}

\title{Fractional nonlinear surface impurity in a 2D lattice}

\author{Mario I. Molina}
\affiliation{Departamento de F\'{\i}sica, Facultad de Ciencias, Universidad de Chile, Casilla 653, Santiago, Chile}

\date{\today }

\begin{abstract} 
We study the formation of localized modes around a generalized nonlinear impurity which is located at the boundary of a semi-infinite square lattice, and where we replace the standard discrete Laplacian by a fractional one, characterized by a fractional exponent $0<\alpha<1$ where $\alpha=1$ marks the standard, non-fractional case. We specialize to two impurity cases: impurity at an ``edge'' and impurity at a ``corner'' and use the formalism of lattice Green functions to obtain in closed form  the bound state energy and its mode amplitude.  It is found that, for any fractional exponent and for impurity strengths above a certain threshold, there is always a single bound state for the linear impurity, while for the nonlinear (cubic) case, up to two bound states are possible.  At small fractional exponents, the energy of the impurity mode becomes directly proportional to the impurity strength. 
\end{abstract}

\maketitle

{\bf 1.\ Introduction}.\\ The concept of fractional calculus and its ever-increasing applications to physics is a growing area. It all started in $1695$ from  correspondence between Leibniz and L'Hopital who discussed the possibility of defining non-integer  derivatives. For instance, they wondered, `what is the half-derivative of a function'? A quick initial answer can be found as follows:  Given a rather general function $f(x)$ expressed a a power series $f(x)=\sum_{n} c_{n} x^n$, the relevant derivative is that of $x^n$. This derivative, in turn, can be expressed in terms of the gamma function: The $\alpha$th derivative of $x^n$ can be expressed as
$(d^\alpha/d x^\alpha) x^n = ( \Gamma(n+1)/\Gamma(n-\alpha+1) )\ x^{n-\alpha}$
which is well-defined for any real $\alpha$. By using this result, the fractional derivative of $f(x)$ can be computed in principle by deriving the series term by term. There are ambiguities with this simple procedure, however. For instance, the $\alpha$th derivative of a constant function $f(x)=1$ is not necessarily zero. Direct application of the previous prescription gives $(d^\alpha/d x^\alpha)=(1/\Gamma(1-\alpha)) x^{-\alpha} \neq 0$. However, one could have done $(d^{\alpha-1}/d x^{\alpha-1})(d/dx)\ 1 = 0$. This brings into question the commutativity and additivity of the derivative operator. The mathematical efforts to bring formal consistency to these ideas, have promoted fractional calculus from a mathematical curiosity to a full-fledged research field\cite{fractional1,fractional2,fractional3,hilfer}.
Several mathematically sound definitions for the fractional derivative have been given over the years, each one with its advantages and disadvantages. two of the most used definitions are the Riemann-Liouville form
\be
\left({d^{\alpha}\over{d z^{\alpha}}}\right) f(z) = {1\over{\Gamma(1-\alpha)}} {d\over{d z}} \int_{0}^{z} {f(x)\over{(z-x)^{\alpha}}} dx,
\ee
and the Caputo formula
\be
\left({d^{\alpha}\over{d z^{\alpha}}}\right) f(z) = {1\over{\Gamma(1-\alpha)}} \int_{0}^{z} {f'(x)\over{(z-x)^{\alpha}}} dx
\ee
where, $0<\alpha<1$. This unorthodox formalism with its definitions of a fractional integral and fractional derivative, has found application in several fields: fluid mechanics\cite{fluid2}, fractional kinetics and anomalous diffusion\cite{metzler,sokolov,zaslavsky}, optics\cite{optica1,optica2,optica3,optica4,optica5,optica6}, strange kinetics\cite{shlesinger}, Levy processes in quantum mechanics\cite{levy}, fractional quantum mechanics\cite{laskin,laskin2}, plasmas\cite{plasmas}, electrical propagation in cardiac tissue\cite{cardiac}, biological invasions\cite{invasion}, and epidemics\cite{epidemics}. It has proven most useful in describing the behavior of systems with nonlocal interactions and systems with memory effects.

In this work, we examine the effects of this fractionality on a simple system consisting of an isolated impurity embedded in a 2D periodic lattice, where the impurity is taken as linear or nonlinear (cubic). Single impurity problems are appealing since they serve as a bridge between ordered and fully disordered systems. They are relatively simple to solve and sometimes an exact, closed-form solution can be obtained.
In the absence of fractionality and nonlinearity, it is well-known that the periodic lattice possesses a band spectrum with plane wave eigenmodes. The breaking of the discrete translational invariance caused by the presence of a linear impurity causes that 
the state at the band edge detaches and gives rise to a localized state centered around
the impurity position. The rest of the modes become somewhat perturbed by the presence of the impurity but retain their extended character. It has been proven that for 1D and 2D lattices, there is always a localized bound state centered at the impurity\cite{slater,harrison}, no matter how weak the strength of the impurity.
Some examples of linear impurities include impurity defects and vacancy formation in crystals\cite{impurities1,impurities2}, junction defects between two optical or network arrays\cite{miro}, discrete networks for routing and switching of discrete optical solitons\cite{christo}. They also appear in simple models for magnetic metamaterials, modeled as periodic arrays of split-ring resonators, where magnetic energy can be trapped at impurity positions\cite{wang}. Usually, impurity problems are considered as a first step towards the much harder study of a system containing a finite fraction of impurities, randomly distributed (Anderson). In fact, it has been suggested that there exists an underlying connection between the problem of finding the localized mode in a system with a single impurity, with that of computing the localization of the modes in the Anderson system\cite{econ1,econ2}.

In the case of a nonlinear impurity, this could simply consist of a unit with a nonlinear response immersed in an otherwise linear array of units. For instance, in an inductively-coupled array of electrical units, each consisting of an LC resonator, one can render one of the capacitors as nonlinear by inserting an appropriate nonlinear dielectric between the plates of the capacitor\cite{electric}. In a similar vein, in an optics context, the system of interest is a dielectric waveguide array, where one of the guides is judiciously doped with an element with strong polarizability. 

In previous works\cite{previous1,previous2,previous3,previous4}, we have dealt with the non-fractional standard problem of a general impurity immersed in 1D and 2D lattices, and located at various distances from the boundaries of the lattice. This was accomplished by the use of lattice Green functions, which allow for exact, sometimes closed-form solutions. For the case of a single impurity embedded in a 1D lattice and subject to a fractional Laplacian, it was found that fractionality decreases the impurity threshold needed to create a bound state. Also, the global degeneracy increased with a decrease in fractionality\cite{1D}.

\vspace{0.2cm}
{\bf 2.\ The model}.\\ The starting point is the discrete linear Schrodinger equation containing a single nonlinear impurity, located at position ${\bf d}$ and immersed in a 2D square lattice: 
\be
i {d C_{\bf n}\over{d t}} + 4 C_{\bf n} + \Delta_{n} C_{\bf n} + \chi \delta_{{\bf n},{\bf d}} |C_{\bf n}|^\beta C_{\bf n} = 0,\label{eq2}
\ee
where $\chi$ is the nonlinearity strength, $\beta$ is the nonlinearity exponent and  where $\Delta_{n}$ is the discretized Laplacian:
\begin{eqnarray} 
(\Delta_{n}) C_{\bf n}&=& C_{n_{x}+1,n_{y}}+C_{n_{x}-1,n_{y}}-4\ C_{n_{x},n_{y}}\nonumber\\
                    & &+ C_{n_{x},n_{y}+1}+C_{n_{x},n_{y}-1},
\end{eqnarray}
where ${\bf n}=(n_{x},n_{y})$. We proceed now to replace this discretized Laplacian by its fractional form $(\Delta_{n})^\alpha$, where $(\Delta_{n})^\alpha$ is given by\cite{discrete laplacian,luz}
\be 
(\Delta_{n})^\alpha C_{\bf n} = \sum_{{\bf m}\neq{\bf n} } (C_{\bf m} - C_{\bf n}) \ K^{\alpha}({\bf n}-{\bf m})\label{eq:7}
\ee
where,
\begin{figure*}[t]
 \includegraphics[scale=0.3]{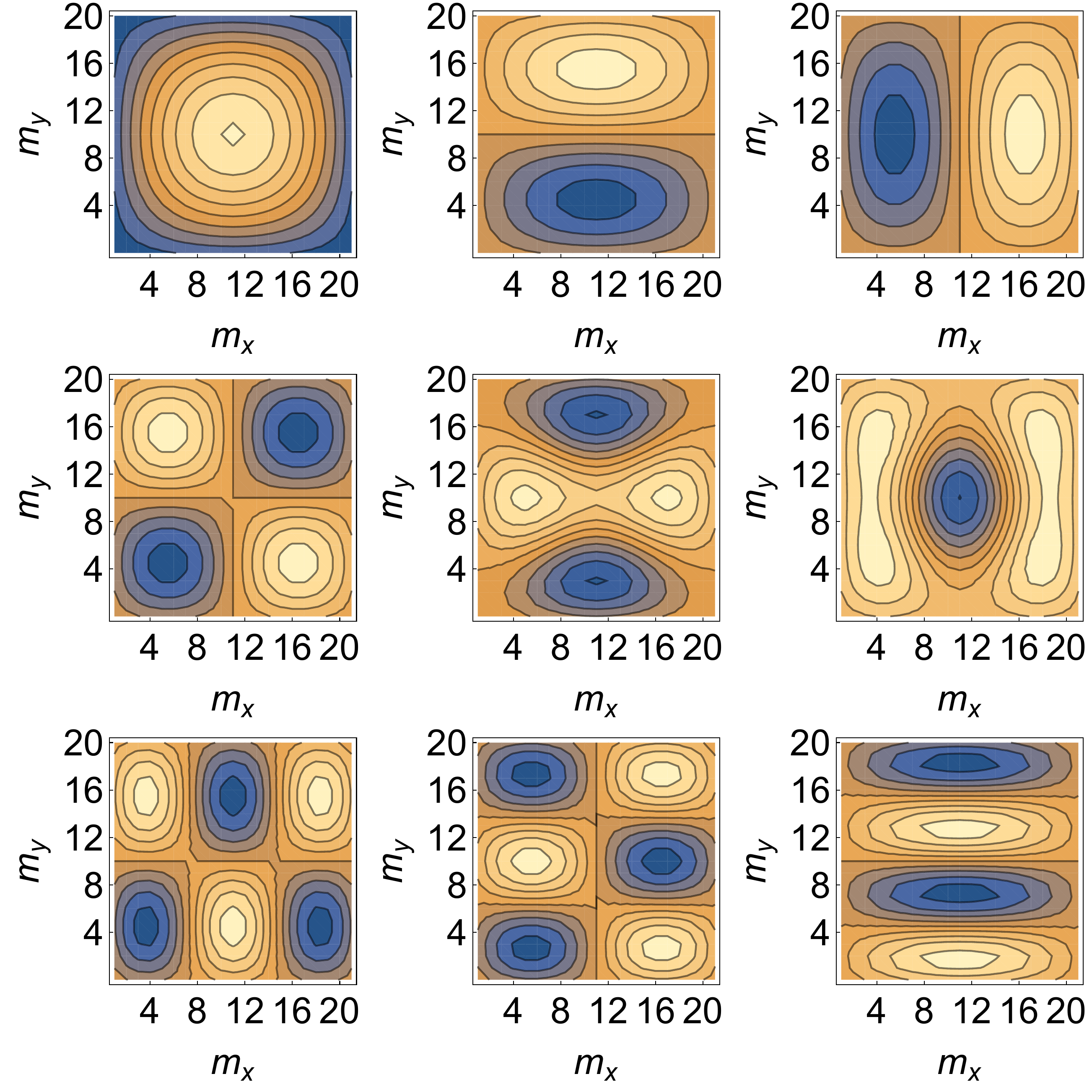}
 \hspace{0.5cm}
 \includegraphics[scale=0.3]{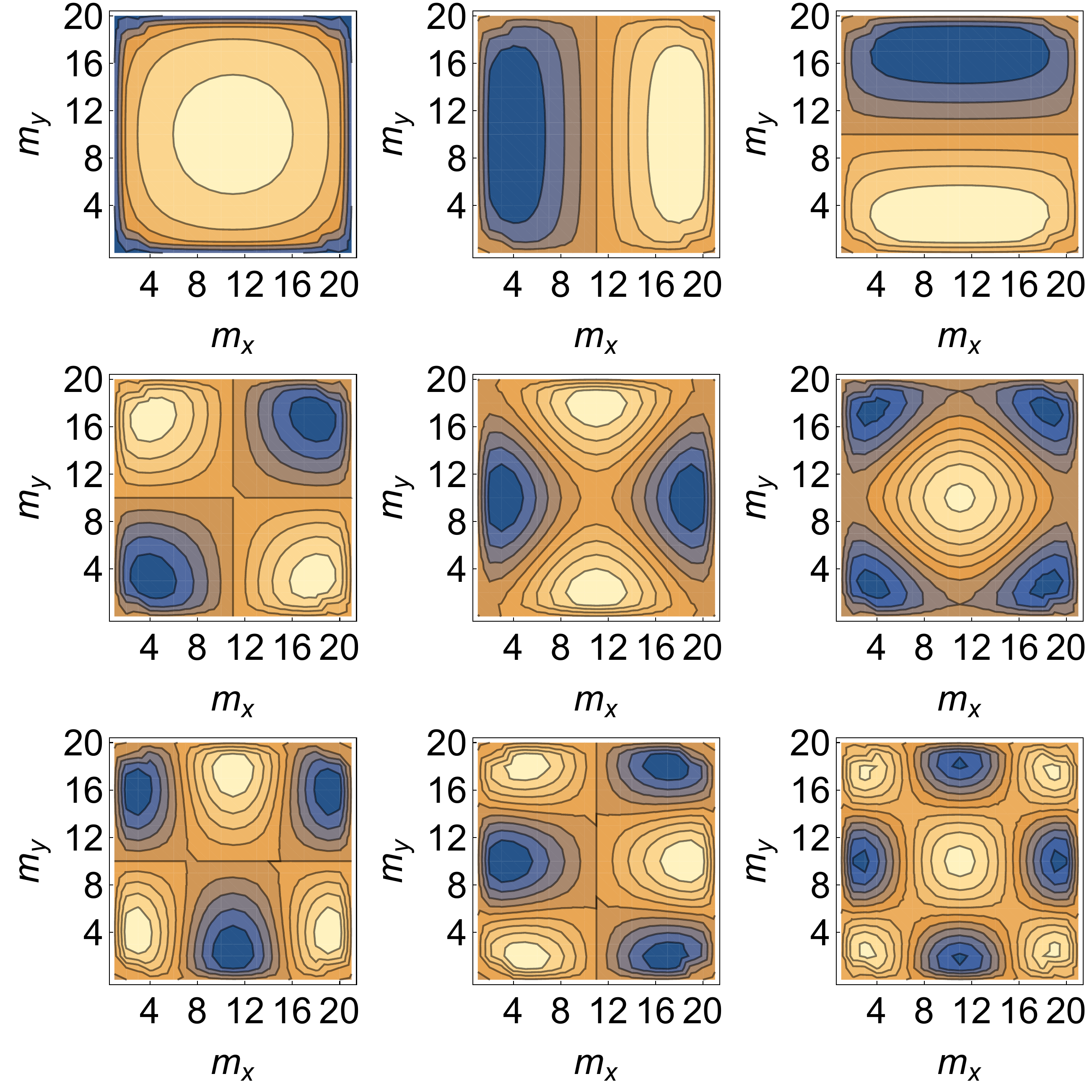}
  \caption{Density plots of the first nine modes for a linear square lattice with fractionality $\alpha=1$ (standard case, left) and $\alpha=0.2$ (right). ($N=21\times 21$) }
  \label{fig1}
\end{figure*}
\be 
K^{\alpha}({\bf n}) = {1\over{|\Gamma(-\alpha)|}}\ \int_{0}^{\infty} e^{-4 t }\ I_{n_x}(2 t)\ I_{n_y}(2 t)\ t^{-1-\alpha}\ dt \label{eq:8}
\ee
with ${\bf n} = (n_x,n_y)$ and $I_{n}(x)$ is the modified special Bessel function.
In the limit $\alpha\rightarrow 1$, $K({\bf m})\rightarrow \delta_{{\bf m},{\bf u}}$ where ${\bf u}=(\pm 1,0)$ or ${\bf u}=(0,\pm 1)$, i,e., coupling to nearest neighbors only, and the system reverts back to Eq.(\ref{eq2}). In the opposite limit, $\alpha\rightarrow 0$, $K({\bf m})\rightarrow 0$.

The stationary modes defined by $C_{\bf n}(t) = e^{i \lambda t} \phi_{\bf n}$ obey\cite{previous3}:
\be
(-\lambda + 4 ) \phi_{\bf n} + \sum_{{\bf m}\neq {\bf n}} (\phi_{\bf m} - \phi_{\bf n}) K^{\alpha} ({\bf m}-{\bf n}) + \chi \delta_{{\bf n},{\bf d}} |\phi_{\bf n}|^\beta \phi_{\bf n} = 0.    \label{stationary}
\ee
For a finite square lattice, the term $4$ is to be replaced by $3\ (2)$ for sites at the edge (corner), in Eq.(\ref{stationary}). 
Let us consider for the moment an infinite lattice with no-impuries, $\chi=0$ and
pose a solution of the type $\phi_{\bf n}= A\ \exp(i {\bf k}.{\bf n})$. Thus, we obtain the dispersion of linear waves\cite{2D}:
\be
\lambda({\bf k}) = 4  + \sum_{\bf n} \left( \cos({\bf k}.{\bf n}) -1 \right)\ K^{\alpha}({\bf n}).
\label{dispersion} 
\ee
Some general properties can be inferred from the behavior of $K^{\alpha}(\bf m)$ and the shape of $\lambda(\bf k)$:\\
(a) For $\alpha\rightarrow 1$, $K^{\alpha}(\bf m)$ is zero, unless ${\bf m}$ is a nearest neighbor of ${\bf 0}$, which implies $\lambda({\bf k})\rightarrow \sum_{nn} \exp(i {\bf k}.{\bf m})$, i.e., the usual tight-binding expression.\\
(b) For $\alpha\rightarrow 0$, $K^{\alpha}(\bf m)$ is zero for all ${\bf m}$, which implies $\lambda({\bf k})\rightarrow 4$. This means a completely flat band, and therefore, complete degeneration of the modes.


Next, let us look at the effect of fractionality on the eigenmodes of Eq.(\ref{stationary}), with $\chi=0$. Results are shown in Fig.1 which compares the first modes for $\alpha=1$ (standard case) with those of $\alpha=0.2$ (well inside the fractional regime). We do not appreciate dramatic differences, although for the fractional case we do observe a tendency for mode  maxima to become wider and shifted towards the system boundaries. In Fig.2 we show the density of states $\Omega(\lambda)=(1/N) \sum_{\bf n} \delta(\lambda - \lambda_{\bf n})$, for several fractional exponents $\alpha$. As the value is decreased from $\alpha=1$ (standard case) to $\alpha=0$, $\Omega(\lambda)$ shifts towards $\lambda=4$, the upper band edge. In the limit $\alpha\rightarrow 0$, $\Omega(\lambda)$ converges to $\delta(\lambda-4)$ signaling complete degeneration.
\begin{figure}[t]
 \includegraphics[scale=0.25]{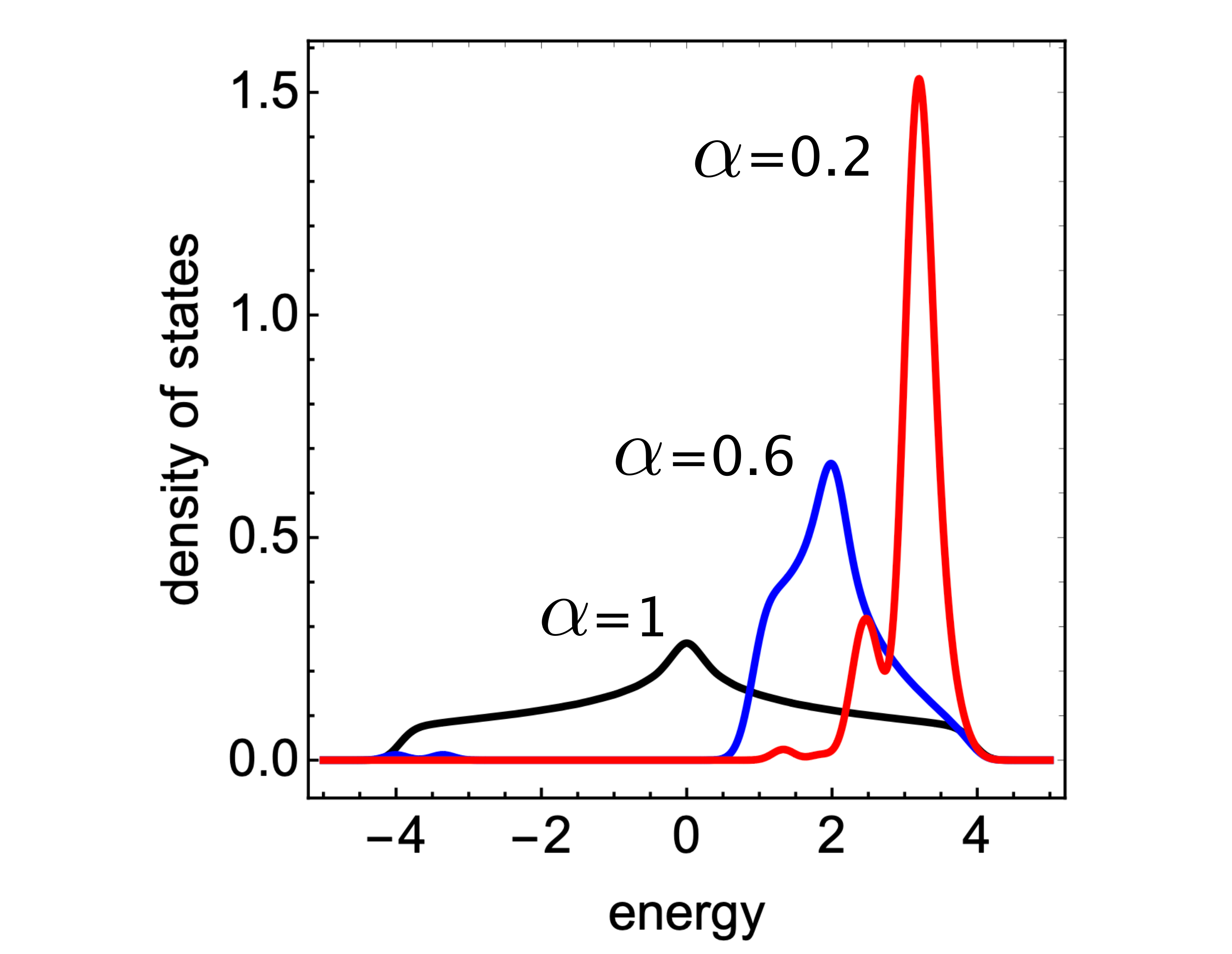}
  \caption{Normalized density of states $\Omega(\lambda)$  for a 2D square lattice with different
  fractional exponents $\alpha$. ($N=21\times 21$) }
  \label{fig2}
\end{figure}

Let us go now to our fractional impurity problem. Consider a single, rather general nonlinear impurity lying on the boundary of a semi-infinite square lattice (Fig.3). We will focus on the existence of a localized mode(s) at the impurity position as a function of fractionality and impurity strength.
\begin{figure}[t]
 \includegraphics[scale=0.17]{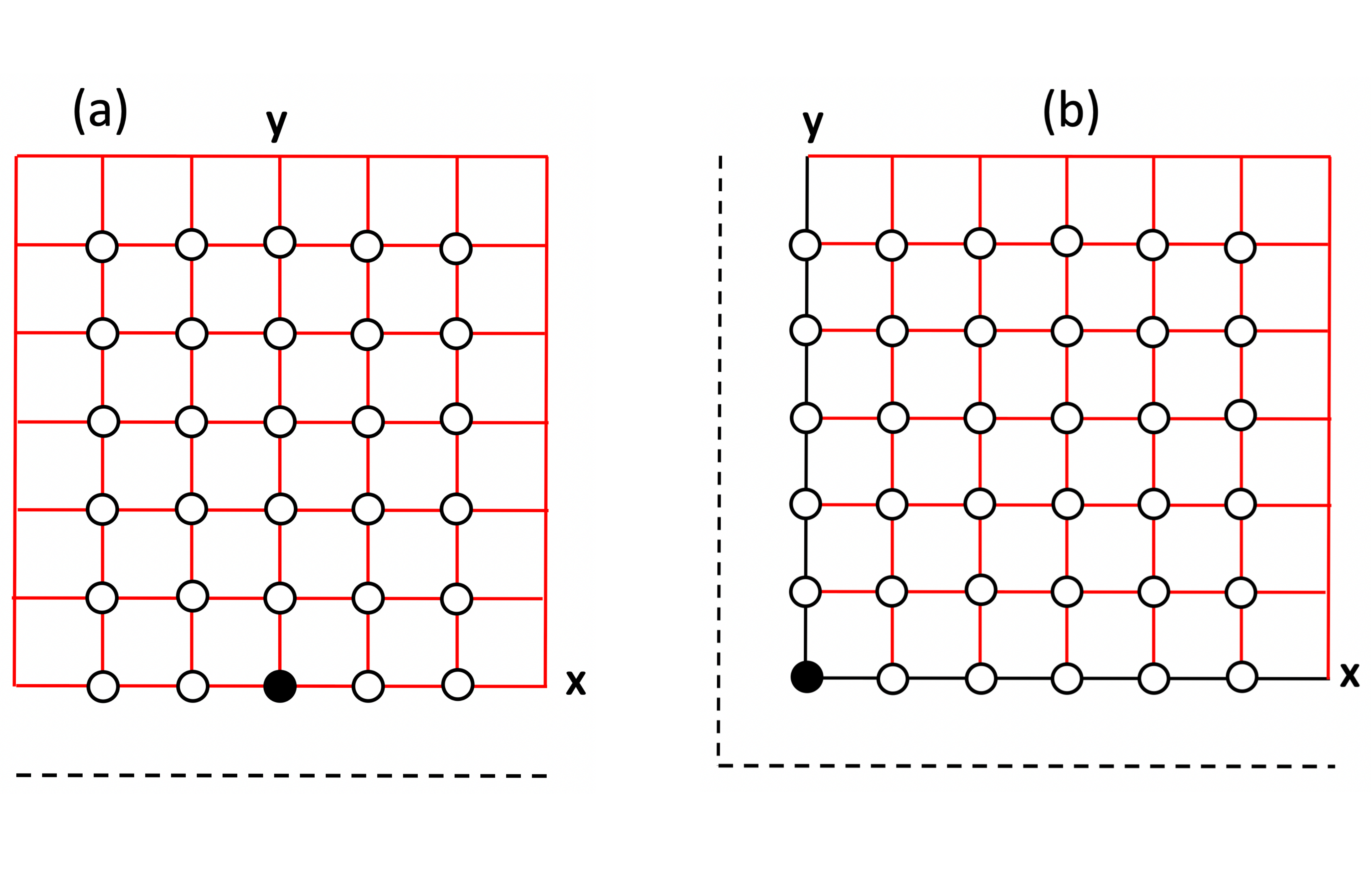}
  \caption{Nonlinear impurity placed at the edge (a) and at the corner (b) of a semi-infinite square lattice. Along the dashed lines the amplitude is strictly zero (After ref.\cite{previous3})}
  \label{fig3}
\end{figure}
The formalism of lattice Green functions is a very elegant and direct method to compute bound states\cite{economou_book}. The lattice Green function is the operator $G = 1/(z-H)$, where $H=H_{0} + V$, and where $H_{0}$ is the unperturbed Hamiltonian and $V$ is the perturbation: $V=\chi |\phi_{\bf d}|^\beta |{\bf d}\rangle\langle {\bf d}|$, where ${\bf d}=(d_{x},d_{y})$ is the position of the impurity and $\phi_{\bf d}$ is the bound state amplitude at position ${\bf d}$.
In this work, we will focus on the cases $\beta=0$ (linear impurity) and $\beta=2$ (nonlinear cubic impurity). The lattice Green function $G$ can be formally expanded in a perturbative series
\be
G=G^{(0)}+ G^{(0)} V G^{(0)}+G^{(0)} V G^{(0)} V G^{(0)}+\cdots.
\ee
For our perturbation $V$ , the series can be resumed to all orders to yield\cite{previous3}
\be
G_{\bf m n} = G^{(0)}_{\bf m n} + {\chi |\phi_{\bf d}|^\beta G^{(0)}_{\bf m d}G^{(0)}_{\bf d n}\over{1-\chi|\phi_{\bf d}|^\beta G^{(0)}_{\bf d d}}}
\ee
where $G_{\bf m n}\equiv \langle {\bf m}|G|{\bf n}\rangle$. The energies of the bound state $E_{b}$ are obtained from the pole(s) of $G_{\bf m n}$, i.e,
\be
1 = \chi |\phi_{\bf d}|^\beta G^{(0)}_{\bf d d}(E_{b}).\label{poles}
\ee
On the other hand, the bound state amplitudes $\phi_{\bf n}$ are obtained from the residues of $G_{\bf m n}$ at $E=E_{b}$\cite{previous3}:
\be
|\phi_{\bf d}|^2 = - {G^{(0)}_{\bf n d}(E_{b}) G^{(0)}_{\bf d n}(E_{b})\over{G'^{(0)}_{\bf d d }(E_{b})}}.\label{residues}
\ee
Inserting Eq.(\ref{residues}) into Eq.(\ref{poles}) leads to a nonlinear equation for the bound state energies\cite{previous3}:
\be
{1\over{\chi }} = {{G^{(0)}}^{\beta+1}_{\bf d d}(E_{b})\over{[-G'^{(0)}_{\bf d d}(E_{b})]^{\beta/2}}}.\label{poles2}
\ee
The unperturbed Green function $G^{(0)}_{\bf m n}(z)$ is given by\cite{economou_book}
\be
G^{(0)}_{\bf m n}(z) = {1\over{(2 \pi)^2}} \int_{-\pi}^{\pi}\int_{-\pi}^{\pi} {\exp(i {\bf k}.{({\bf n-m})}) d^2 k\over{z - \lambda({\bf k})}}.\label{18}
\ee
In our case, $\lambda({\bf k})$ is the dispersion given by Eq.(\ref{dispersion}) and because of its complexity, it is not possible to obtain a closed form expression for $G^{(0)}_{\bf m n}(z)$. In what follows we compute 
$G^{(0)}_{\bf m n}(z)$ numerically, in order to solve Eqs.(\ref{poles2}),(\ref{residues}) for the bound state energies and bound state amplitudes.
Now, it turns out that $G^{(0)}_{\bf m n}$ is not quite the Green function we need. Because our impurity lies on the boundary of a semi-infinite lattice, 
`image' effects must be taken into account as shown next.

\vspace{0.2cm}
{\bf 3.\ Results}.\\
 
{\em Edge impurity}. Let us consider an impurity located at distance $d$ from the edge, ${\bf d}=(0,d)$, where we will take $d=0$ at the end. Since there is no lattice below $(0,0)$,  $G^{(0)}_{\bf m n}(z)$ 
should vanish identically along the sites lying on the dashed line in Fig.2a. 
This means that the actual unperturbed Green function should be
\be
G^{(0)}_{\bf d d}=G^{(\infty)}_{\bf d d} - G^{(\infty)}_{{\bf d},{-{\bf d}-2{\bf j}}}
\ee
where ${\bf j}$ is a unit vector in the y direction and where $G^{(\infty)}_{\bf m n}$ refers to the Green function of the infinite 2D square lattice, Eq.(\ref{18}). 
Now, using the properties $G^{(\infty)}_{\bf m n}=G^{(\infty)}_{\bf n m}=
G^{(\infty)}_{\bf m-n} $, we can write in a simplified notation 
\be
G^{(0)}(0,0;z) = G^{\infty}(0,0;z)-G^{\infty}(0,2 d+2;z)\label{eq20}
\ee
where $G^{\infty}(m,n;z)$ refers to the Green function for an infinite square lattice. At $d=-1$
(dashed line in Fig.2b), we have $G^{(0)}=0$. When the impurity is right at the edge, $d=0$ and 
\be
G^{(0)}(0,0;z) = G^{\infty}(0,0;z)-G^{\infty}(0,2;z)\label{eq21}
\ee

{\em Corner impurity}.\ In this case, the number of images increase to three. For simplicity let us approach the corner site along a diagonal, ${\bf d}=(d,d)$. Now, there is no lattice to the left or below $(0,0)$, so $G^{(0)}_{\bf m n}$ should vanish along the dashed line in Fig.3b. This constraint implies
\begin{eqnarray}
G^{(0)}_{\bf d d}(z)&=&G^{\infty}_{\bf d d} - G^{\infty}_{{\bf d},(d_{x},-d_{y}-2)} - G^{\infty}_{{\bf d},(-d_{x}-2,d_{y})} + \nonumber\\
                    &  & +\  G^{\infty}_{{\bf d},(-d_{x}-2,-d_{y}-2)}
\end{eqnarray}
or, in a simplified notation,
\be
G^{(0)}_{\bf d d} = G(0,0;z)-2 G(0,2d+2;z)+G(2d+2,2 d+2;z).
\ee
When the impurity lies at the corner, $d=0$ and 
\begin{eqnarray}
G^{(0)}(0,0;z)&=& G^{\infty}(0,0;z)-2 G^{\infty}(0,2;z) + \nonumber\\
               & & +\  G^{\infty}(2,2;z)\label{eq24}
\end{eqnarray}

For both cases, edge and corner impurities, the equation for the bound state energy (\ref{poles2}) can be written in the simplified notation as
\be
{1\over{\chi }} = {{G^{(0)}}^{\beta+1}(0,0;z_{b})\over{[-G'^{(0)} (0,0,z_{b})]^{\beta/2}}},\label{poles3}
\ee
where $G^{(0)}$ is given by Eq.(\ref{eq21})[(\ref{eq24})] for the edge (corner) impurity. Finally, the unperturbed Green function for the bulk, Eq.(\ref{18}) can be written as
\be
G^{(0)}(m_{x}, m_{y}; z)= {1\over{(2 \pi)^2}} \int_{-\pi}^{\pi}\int_{-\pi}^{\pi} {\exp(i {\bf k}.{{\bf m}})\ d^2 k\over{z - \lambda({\bf k})}}.\label{25}
\ee
We consider now a linear ($\beta=0$) impurity and a nonlinear  (cubic, $\beta=2$) one.
\begin{figure}[t]
 \includegraphics[scale=0.2]{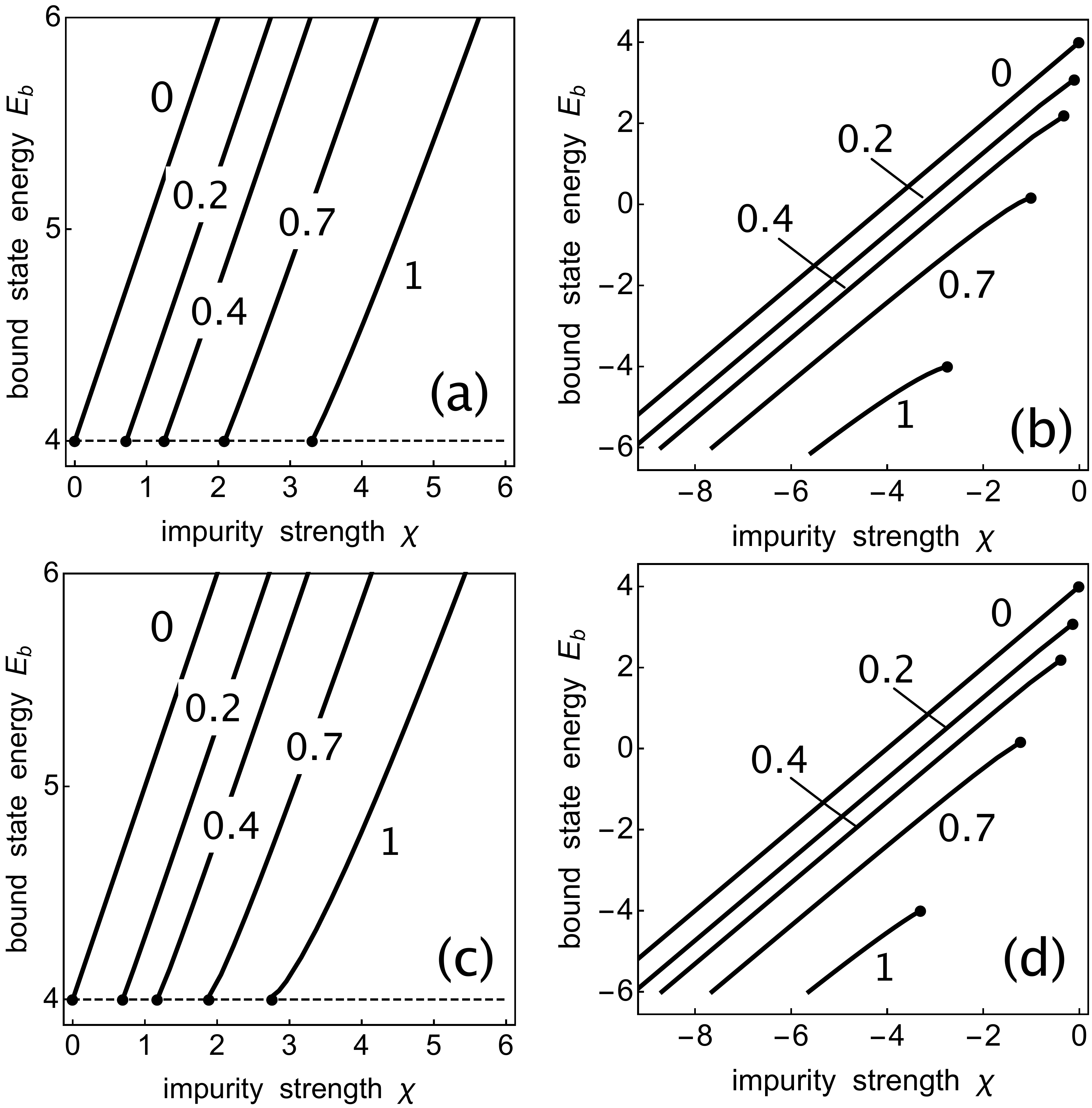}
  \caption{Bound state energies $E_{b}$ for a {\bf linear} impurity versus impurity strength $\chi$, for positive (left column) and negative (right column) impurity strengths, for an impurity at the edge (a,b) and at the corner (c,d) of a semi-infinite square lattice. The numbers on each curve denote the value of the fractional exponent $\alpha$ ($N=11\times 11$).}
  \label{fig4}
\end{figure}
\begin{figure}[t]
 \includegraphics[scale=0.2]{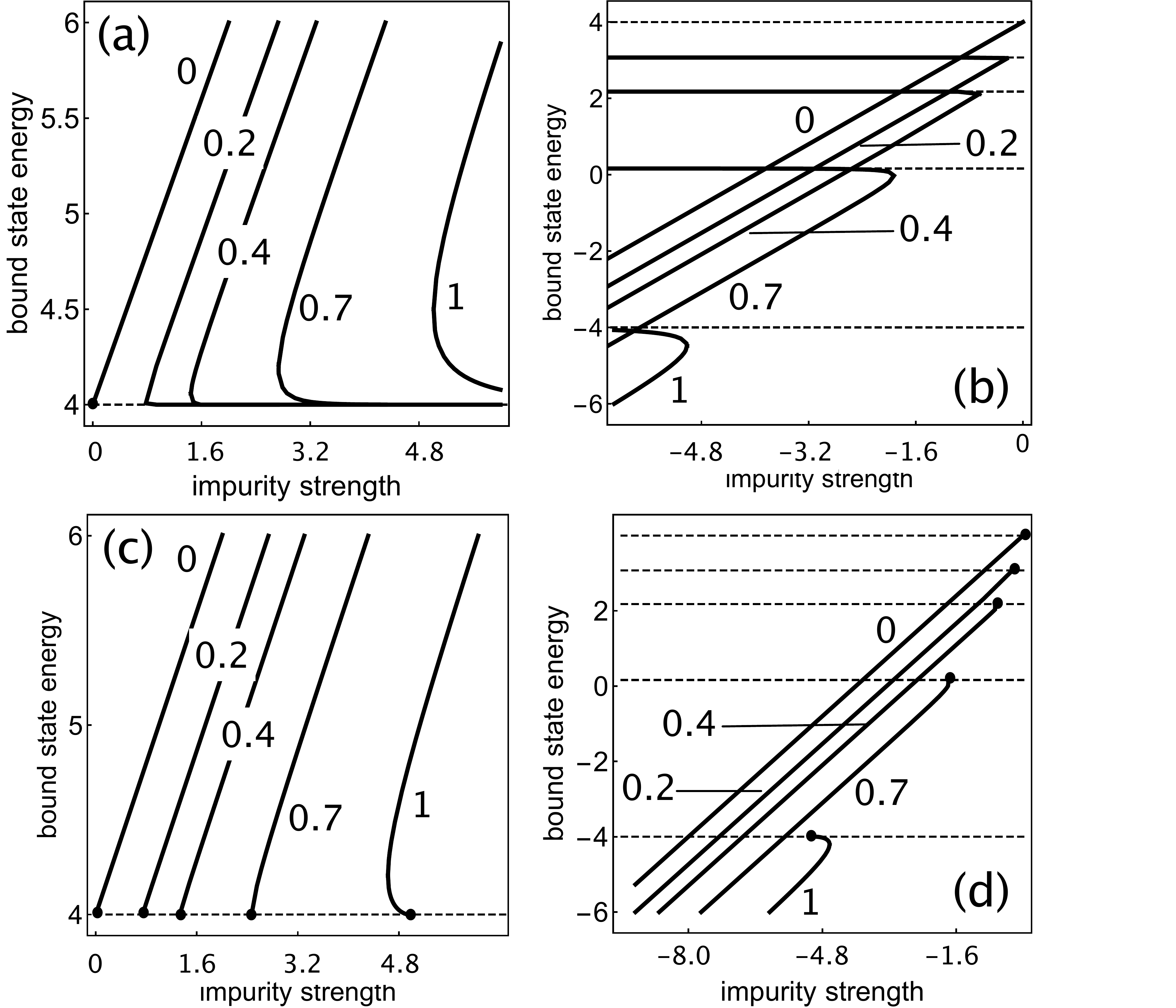}
  \caption{Bound state energies $E_{b}$ for {\bf nonlinear} (cubic) impurity versus  impurity strength $\chi$, for positive (left column) and negative (right column) impurity strengths, for an impurity at the edge (a,b) and at the corner (c,d) of a semi-infinite square lattice. The numbers on each curve denote the value of the fractional exponent $\alpha$ ($N=11\times 11$).}
  \label{fig5}
\end{figure}
For the linear surface impurity, results are shown in Fig.4. We note that, for all values of the fractional exponent $\alpha$, there is a minimum value for the magnitude of the impurity strength $|\chi|$ to effect a bound state. This is true for all $\alpha$, ranging from $\alpha=1$ (standard case) down to $\alpha=0$. The magnitude of the minimum $|\chi|$ decreases with $\alpha$, being the highest for $\alpha=1$ and $0$ for $\alpha=0$, as Fig.4 shows. 

We also note that the upper band edge does not change with $\alpha$, but the lower end does, with lower band edges at positions: $-4 (\alpha=1)$, $0.1629 (\alpha=0.7)$,
$2.1782 (\alpha=0.4)$, $3.0713 (\alpha=0.2)$ and $4 (\alpha=0)$. Thus, at negative impurity strengths, we have a steady decrease of the bandwidth, accompanied by a decrease in  the amount of impurity needed to support a bound state.

For the case of a nonlinear (cubic) impurity (Fig.5), the situation is kind of similar, except that now there can be two bound states for a given impurity strength. The effect is most noticeable for $\alpha=1$. Of the two modes, one of them approaches the band edge, while the other detaches from it, as the magnitude of the impurity strength increases. As argued before in previous works\cite{seis,siete} the former should correspond to an unstable localized state, while the latter denotes a stable bound state. 
The state that approaches the band is created very close to it, so it is sometimes hard to distinguish it from the band (at the scale used). 
As in the case of the linear impurity, there is also a minimum $|\chi|$ to create a bound state. The impurity strength needed to create a bound state decreases with a decrease in the fractional exponent. In fact, the standard, non-fractional case ($\alpha=1$) is the one with the highest impurity threshold.

For both positive and negative impurity strengths, the eigenvalue equation that gives rise to the bound states is qualitatively similar and the only relevant difference is that, at negative impurity strengths, the lower band edge approaches the upper edge as $\alpha$ is decreased, affecting in this way the numerical values of the bound state energy. The presence of $\alpha$ breaks the symmetry that is present in the standard case ($\alpha=1$) where, for $\chi>0 (<0)$, one has $z_{b}>0 (<0)$. For $\alpha\neq 1$ the eigenvalue asymmetry is evident in Figs. 4 and 5.

At $\alpha=0$, the energy of the lower edge becomes exactly $4$, and the band becomes completely flat with an energy curve that is straight function of $\chi$. This is easy to explain on general grounds:
For $\alpha\rightarrow 0$, $\lambda\rightarrow 4$, which implies $G^{(0)}_{\bf m n}(z)\rightarrow (1/(z-4))\ \delta_{{\bf m},{\bf n}}$. Thus, when solving the energy equation (\ref{poles3}), one obtains $1/\chi = 1/(E_{b} - 4)$, i.e., $E_{b}=\chi + 4$, valid for any $\beta$. This 
also implies $|\phi^{b}_{\bf n}|^2 \rightarrow \delta_{\bf n,0}$, a completely localized mode.

\vspace{0.3cm}

{\bf 4.\ Conclusions}\\ \\
In this work, we have re-examined the old problem of a tight-binding nonlinear impurity on a square lattice when we consider fractional effects. The fractional Laplacian gives rise to an effective coupling whose range depends on the value of the fractional exponent $\alpha$. In the absence of impurity, the dispersion of the plane wave modes shows a narrowing of the bandwidth as the exponent decreases; the mean square displacement is always ballistic, with a speed that decreases steadily with a decrease of the fractional exponent. Also, the density of states shows a tendency towards increasing degeneracy as the exponent decreases away from $1$ (the non-fractional case). In the presence of the impurity, we resorted to the formalism of lattice Green functions. Here, the only formal change from the non-fractional case, is that now the energy spectrum depends on the fractional exponent. The rest of the formalism including how to compute the bound state energy and the computation of the mode amplitude is the same as before. Because of the complexity of the resulting dispersion, it is not possible to obtain closed-form expressions for the Green functions. In addition, since the impurity lies along the boundary of the semi-infinite square lattice, we had to take into account image effects where the relevant Green function is a superposition of standard Green functions used for the infinite lattice.
Due to the complexity of the system, we had to compute the relevant Green functions numerically to arrive at the bound state energies as a function of the impurity strength, for several fractional exponents.  For both cases, linear and nonlinear impurity, the main effect of fractionality was to shift the minimum impurity strength needed for a bound state to exist steadily towards smaller values as the exponent decreased. In the limit of a vanishing exponent, the energy curve converged to a straight line, becoming directly proportional to the impurity strength $\chi$. This result actually holds for any nonlinearity exponent $\beta$. This phenomenology agree with the results found for the 1D case\cite{1D}. All in all, the square lattice with a fractional defect is not dramatically different from its non-fractional version, and its main features are similar to what was found before for a one-dimensional lattice.  Thus, the phenomenology of a system with a single impurity seems  robust against a fractional extension of the Laplacian operator, which is interesting in itself.

As mentioned in the Introduction, it has been suggested that there is a formal correspondence between the problem of Anderson localization and that of the existence of a bound state for the single bulk linear impurity\cite{econ1,econ2}.  Thus, for $D=1,2$ there is always a bound state regardless of the impurity strength, implying that  all states are localized for the corresponding Anderson system. For $D=3$, a minimum impurity strength is needed to create a bound state, which implies that a minimum disorder strength is needed for Anderson localization to occur.   
While the above correspondence has been suggested for a single bulk, linear impurity, it would be interesting to explore whether this correspondence will also hold for  a more general type of impurity (boundary and/or nonlinear and/or fractional). If so,  this correspondence mentioned above would imply concrete predictions for the Anderson version of our system. From Figs.4 and 5 we see that for all fractional exponent values, a minimum impurity strength must be reached for a bound state to exist both, for the linear and nonlinear case. Following the above correspondence, this would imply that, for a semi-infinite, linear/nonlinear and fractional Anderson system, a minimum disorder strength would be needed for localization. It is interesting to note a previous experimental result that seem to partially confirm the above picture. The system is a  1D truncated disordered photonic  array,  where it was measured that a higher level of disorder near the boundary of the array is needed to obtain a similar degree of localization as that at the bulk\cite{szameit}. We intend to delve into these interesting issues in a forthcoming work.

\vspace{0.5cm}
{\bf Acknowledgments}

This work was supported by Fondecyt Grant 1200120.

\end{document}